\documentclass[preprint,showkeys]{revtex4}
\usepackage{amssymb,amsmath}
\usepackage{epsfig}
\font\bba=msbm10 scaled 1080
\font\bbb=msbm8 
\font\bbc=msbm6 
\newfam\bbfam
\textfont\bbfam=\bba
\scriptfont\bbfam=\bbb
\scriptscriptfont\bbfam=\bbc
\def\bb{\fam\bbfam\bba}

\def\Z{{\bb Z}}

\begin{document}
\title{Liquid-Vapor Transition and Critical Behavior of The Ultrasoft Restricted
Primitive Model
of Polyelectrolytes : a Monte Carlo Study}
\author{J.-M. Caillol}
\email{Jean-Michel.Caillol@th.u-psud.fr}
\author{D. Levesque}
\email{Dominique.Levesque@th.u-psud.fr}
\affiliation{Laboratoire de Physique Th\'eorique, UMR 8627, b\^atiment 210}      
\affiliation{Univ. Paris-Sud and CNRS, Orsay, F-91405, France}                                                         
\date{\today}        
\begin{abstract}
We present a Monte-Carlo study of the liquid-vapor transition and the critical behavior
of a model of polyelectrolytes
with soft  gaussian charge distributions introduced recently by  Coslovich,  Hansen,  and  Kahl 
[J. Chem. Phys.  \textbf{134}, 244514 (2011)].
A finite size study involving four different volumes in the grand canonical ensemble yields 
a precise determination of the critical temperature, chemical potential, and density of the model.
Attempts to determine the nature of the criticality and to obtain reliable values for the critical
exponents are not conclusive.
\end{abstract}
\keywords{Coulomb criticality; Monte Carlo simulations;  Grand canonical ensemble; Critical phenomena;  Finite size effects}
\maketitle                                                                                
\section{Introduction}
\label{intro}
After nearly twenty years of endeavors to elucidate the nature of the critical behavior of
classical Coulombic liquids, no firm conclusion concerning this issue has yet been provided by theory, experiment 
and computer simulation. 
The long range of Coulomb potential would suggest classical mean field criticality, while
the Debye-H\"{u}ckel screening,  yielding  short range effective interactions, would rather suggest
an Ising-like critical behavior \cite{Fisher-I,Stell-I,Fisher-II,Stell-II,Stell-III,Wein}.
The more recent contributions of the Orsay group on the type of Coulomb criticality is a finite size scaling (FSS)
analysis of
grand canonical Monte Carlo simulations of the liquid-vapor transition of the three dimensional (3D)
restricted primitive model of electrolytes (RPM), \textit{i.e.}~a fluid of charged hard spheres of
opposite charges and equal diameters
\cite{Caillol-I}. The conclusions of this analysis unambiguously discard mean field behavior and rather suggest that
the 3D-RPM belongs to the universality class of the 3D-Ising model.  After a lot of controversy, most
authors who contributed to this issue seem now in  favor of  an Ising-like criticality~\cite{Luij-Fish,Pana}.


The numerical simulations of the RPM in the critical region  are notably plagued by many numerical difficulties :
the usual critical slowing down and, more specifically, the
long range of the Coulombic interaction and the unusually low  values (in natural reduced units)
of the  critical temperature ($T_{c}^{*}=0.049 17(2)$) and the density ($\rho_{c}^{*}=0.080(5)$).
Therefore the new model of polyelectrolytes introduced in  Ref.~\cite{Coslo-I,Coslo-II,Nikou} is welcome and could well
provide a new interesting toy model for studying Coulomb criticality, irrespective of its validity to  reproduce
 the physics of polyelectrolytes. 
The ultrasoft restricted primitive model
(URPM) of polyionic solutions is a mixture of positive and negative extended  charge distributions, hence the alternative denomination
of ''fuzzy polyelectrolytes''. To make things easier all polyions  share the same shape factor $\tau( \mathbf{r})$. Of course the net global
charges of cations and anions are of opposite values $\pm Q$. The system is H-stable in the sense of 
Fisher and Ruelle \cite{Coslo-I,Ruelle, Fisher-Ruelle} and therefore
admits a well behaved thermodynamic limit. The key point is that additional short range interactions such hard cores
are not required to ensure thermodynamic stability.
Of course both the RPM and the URPM are expected to belong to the same universality class.  The URPM is conceptually
the simplest model of Coulombic fluid with -possibly- a critical point and it thus deserves the closest attention.

The authors of  Ref.~\cite{Coslo-I,Coslo-II,Nikou} made the choice of a gaussian
distribution for  $\tau( \mathbf{r})$ which   yields simple  analytical expressions for the ionic interactions.
Their Monte Carlo and Molecular Dynamics simulations of the gaussian  URPM  in the fluid phase reveal
the existence of 
a liquid-vapor coexistence curve. 
It is noteworthy that the many theories examined in Refs~\cite{Nikou,Warren-I,Warren-II,Warren-III} seem
 to be unable to reproduce the simulation
results in the vicinity of the critical point.

Here we report new MC simulations in the critical region aimed at a more
precise location of the critical point and an attempt to obtain  the critical exponents.
These simulations were performed
in the grand-canonical (GC) ensemble using a cubic simulation cell with periodical boundary 
conditions together with  Ewald potentials ~\cite{Frenkel}.
We considered  4 different volumes $V=L^3$ and, for each volume, we  made use of histogram reweighting 
to determine the histograms $p_L(\rho,u)$  ($\rho$ numerical density, $u$ energy per unit volume)
in a domain of temperatures and chemical potentials close to that of 
the critical point~\cite{Ferrenberg}. 

Our paper is organized as follows: in section~\ref{model} we explicit the model 
and we then  give details on our simulations in section~\ref{MC}. The results
are discussed in section~\ref{Results} and conclusions are drawn in section~\ref{Conclusion}.

\section{The model}
\label{model}
The URPM is an equimolar mixture of $N_{+}=N/2$ cations of charge $+Q$ and $N_{-}=N/2$ anions
of charge $-Q$ in a volume $V$. Cations and anions bear an extended charge distribution $\pm Q  \tau( \mathbf{r})$ 
where the distribution  $ \tau( \mathbf{r})$, normalized to unity, is supposed to be the same for all 
species of polyions and given by a Gaussian law : 
\begin{equation}
\label{distribution}
 \tau( \mathbf{r})=\left(\dfrac{1}{2\pi \overline{\sigma}^{2}}\right)^{3/2}\; \exp(-\mathbf{r}^2/(2  \overline{\sigma}^{2}) ) \; ,
\end{equation}
where $\overline{\sigma} = \sigma/2 $ denotes the radius of the polyion, $\sigma$ its
diameter.
The polyions  interact only through electrostatic interactions.  The pair interaction
between an $\alpha$ and a $\beta$ polyon  \mbox{($\alpha, \beta \; = + \;  \mathrm{ or } \; - $ )}
is given by \cite{Coslo-II}

\begin{equation}
\label{potr}
w_{\alpha \beta}(r) = \dfrac{Q_{\alpha} Q_{\beta}}{r} \; 
\mathrm{erf} (r/ 2 \overline{\sigma})  \; ,
\end{equation}
and, in Fourier space
\begin{equation}
\label{potk}
\widetilde{w}_{\alpha \beta}(k) = \dfrac{4 \pi Q_{\alpha} Q_{\beta}}{k^2} \; 
\exp(- k^2 \overline{\sigma}^2)  \; .
\end{equation}

The polyions  interact only through the electrostatic interactions \eqref{potr} and no additional
soft or hard repulsive interaction is required. In particular, at low temperatures, 
cations and anions   can interpenetrate and form pairs of polarizable dipoles.
We thus expect   the system to behave as  a quasi ideal gas ($\beta p/\rho=1/2$, $p$ pressure)
at low temperatures.
In the case of the usual RPM this possibility is thwarted  by the presence of hard cores.
However, it is easy to show that the configurational
energy of the URPM  is nevertheless bounded by below by a finite  extensive quantity
 $-N_{+}w_{++}(0) -N_{-}w_{--}(0)$ (with
 finite self-energies $u_0 \equiv w_{++}(0)=w_{--}(0) = Q^2/(\sqrt{\pi}  \overline{\sigma})$).
Therefore the system is H-stable in the sense of Fisher and Ruelle
which ensures
the existence of the thermodynamic limit (TL) \cite{Ruelle,Fisher-Ruelle}.  In particular the grand partition function
converges and can be computed without altering its TL value by enforcing the charge neutrality
of all configurations, \textit{i.e.}~imposing $N_{+} = N_{-}$. Indeed a theorem 
by Lieb and Lebowitz ensures that the TL of the unconstrained and neutral  systems are
the same in the GC ensemble~\cite{Lieb}.

To make some contact with the literature on the RPM it seemed  preferable to us to chose
$\sigma$ as the unit of length rather than 
 $\overline{\sigma}$ or $ \sqrt{2}\overline{\sigma}$ as in Refs.~\cite{Coslo-I,Coslo-II,Nikou}.
Henceforth the reduced density of the system is denoted by $\rho^* = N/V^*$ with 
a reduced volume $V^* = V/\sigma^3$. In the same vein we define
the reduced temperature as  $T^*=k_{\mathrm{B}} T / u_0 $
temperature in Kelvin, $k_{\mathrm{B}}$ Boltzmann 
constant) and its dimensionless inverse $\beta = 1/T^*$ (our  unit 
of energy is thus the same as that of Refs.~\cite{Coslo-I,Coslo-II,Nikou}).

\section{Monte Carlo Simulation}
\label{MC}
\subsection{Ewald sums}
\label{geometry}
We considered a cubic simulation cell  of side $L^*=L/\sigma$ with periodic boundary conditions
and we made use of Ewald potentials  to take into account the long range
of Coulomb interactions.
The configurational energy $U$ of the URPM is made of three contributions $U_r$, $U_k$, and $U_s$ of
which the two first are series of functions, respectively defined in direct  and Fourier space,
both with good convergence properties, see \textit{e.g.} \cite{Frenkel}, and $U_s$ is a self-energy term. 
$U$ reads as~\cite{Coslo-II}

\begin{subequations}
\label{Ewald}
\begin{eqnarray}
 U & =& U_r + U_s - U_s \\
\label{r}U_r &=& \frac{1}{2} \; \sum_{i \neq j} \frac{Q_i Q_j}{r_{ij}} \left( \mathrm{erf}(r_{ij}/\sigma) -
                                                                              \mathrm{erf}(r_{ij}/ (\sqrt{2}\widetilde{\sigma}) 
 \right)  \; , \\
\label{k} U_k &=&  \frac{1}{2 V} \sum_{\mathbf{k} \ne \mathbf{0}} \frac{4 \pi}{k^2} \,
                          \exp \left(-\mathbf{k}^2 \widetilde{\sigma}^2/2  \right)
                          | \widetilde{\rho}_{\mathbf{k} }|^2  \; 
\\ 
U_s &=&\frac{N Q^2 }{\sqrt{2 \pi} \widetilde{\sigma} }
\end{eqnarray}
\end{subequations}
In Eqs.~\eqref{Ewald}  $\widetilde{\sigma} = \sqrt{\sigma^{' 2} + \overline{\sigma}^2}$ is related to the 
control parameter $\sigma^{'}$ of the Ewald method. The pair distances $r_{ij}$  in~ \eqref{r} are computed
with the minimum image convention and  the function $\mathrm{erf}(r_{ij})$ is set to zero for
$r_{ij} > L/2$. In Eq.~\eqref{k}
$\mathbf{k}= 2 \pi \mathbf{n}/L $, where $\mathbf{n}  \in \Z^3$ is a  vector  with 3 integer components. 
In practice, only the vectors
with a modulus $||\mathbf{n}|| \leq 7$  are considered in the sum in the r.h.s. of~\eqref{k}. Finally  $ \widetilde{\rho}_{\mathbf{k}}
= \sum_{i=1}^{N} Q_i \exp(i \mathbf{k} \cdot \mathbf{r}_i)$ denotes
the Fourier transform of the microscopic charge density.
In our simulations, in reduced units, $\sigma^{'} =\sigma =1$ and thus  $\widetilde{\sigma}=\sqrt{5}/2$.

\subsection{Grand-Canonical ensemble and Histogram Reweighting}
\label{GC}

\begin{table*}[t!]
\centering
\caption{ \label{Tab-1}   
The table displays, for each reduced volume $V^*$, the range of temperatures $T^*$,
the total number $n_T$ of distinct thermodynamic states $(\mu_i, \beta_i)$, the total number  $n_c$  
of selected configurations (spaced by $1500$ trial moves), the apparent critical temperatures $T_c^*(L) $,
 chemical potentials $\mu_c^*(L) $, and  densities $\rho_c^*(L) $ as they are defined in
Section~\ref{Results}. Numbers in brackets denote the error on the last digit.
}
\begin{tabular}{ || c | c | c | c | c | c | c ||} 
\hline  
 $ V^* $    &      $ T^*  $                                                   &   $n_T $    &   $ n_c$           & $T_c^*(L) $ &   $\mu^*_c(L) $  &   $\rho _c^*(L) $                \\ \hline
 500         &      $ 0.010 \leq T^* \leq    0.019 $               &     $129$    &   $5.3  \, 10^8$         &  0.0181(1)  &  -0.192104(1)  &   0.21(1)        \\   \hline 
 1000       &      $ 0.011 \leq T^* \leq    0.019 $               &     $122$    &   $5.7   \, 10^8$        &  0.0162(1) &   -0.189923(1) &    0.225(5)    \\  \hline 
 2000       &      $ 0.012 \leq T^* \leq    0.019 $               &     $181$    &   $1.1  \, 10^{9}$       &  0.0150(1) &   -0.188656(1) &   0.23(1)    \\  \hline 
 4000       &      $ 0.013 \leq T^* \leq    0.018 $               &    $59$       &   $1.8  \, 10^{9}$       &  0.0142(1)  &  -0.187934(1) &     0.27(1)  \\  \hline 
\end{tabular}
\end{table*}
We performed  MC simulations in the GC  ensemble which is well suited for the simulation
of multiphase systems \cite{Frenkel}.
In the GC ensemble the volume $V$, the inverse temperature $\beta $
and the chemical potential $\mu $ are fixed.
We considered 4 different reduced volumes $V^*= 500, 1000, 2000$ and $4000$
in order to study and exploit FSS effects.
This should be compared with the simulations of Ref.~\cite{Coslo-I}, also performed in the grand canonical
ensemble for volumes of  $V^*= 282$ and $V^*=1123$ in our units.
It was not necessary to use the biased MC schemes usually considered for the RPM (see e. g. Ref.~\cite{Caillol-I,Pana})
and we thus made use of the standard Metropolis algorithm~\cite{Frenkel}. In addition to the trial displacements of individual polyions
we only considered trial  insertions or deletions of a single  pair of  an anion plus a cation in order to preserve
the overall charge neutrality $N_+=N_-$.
\begin{figure}[t!]
\centering
\includegraphics[angle=0,scale=0.60]{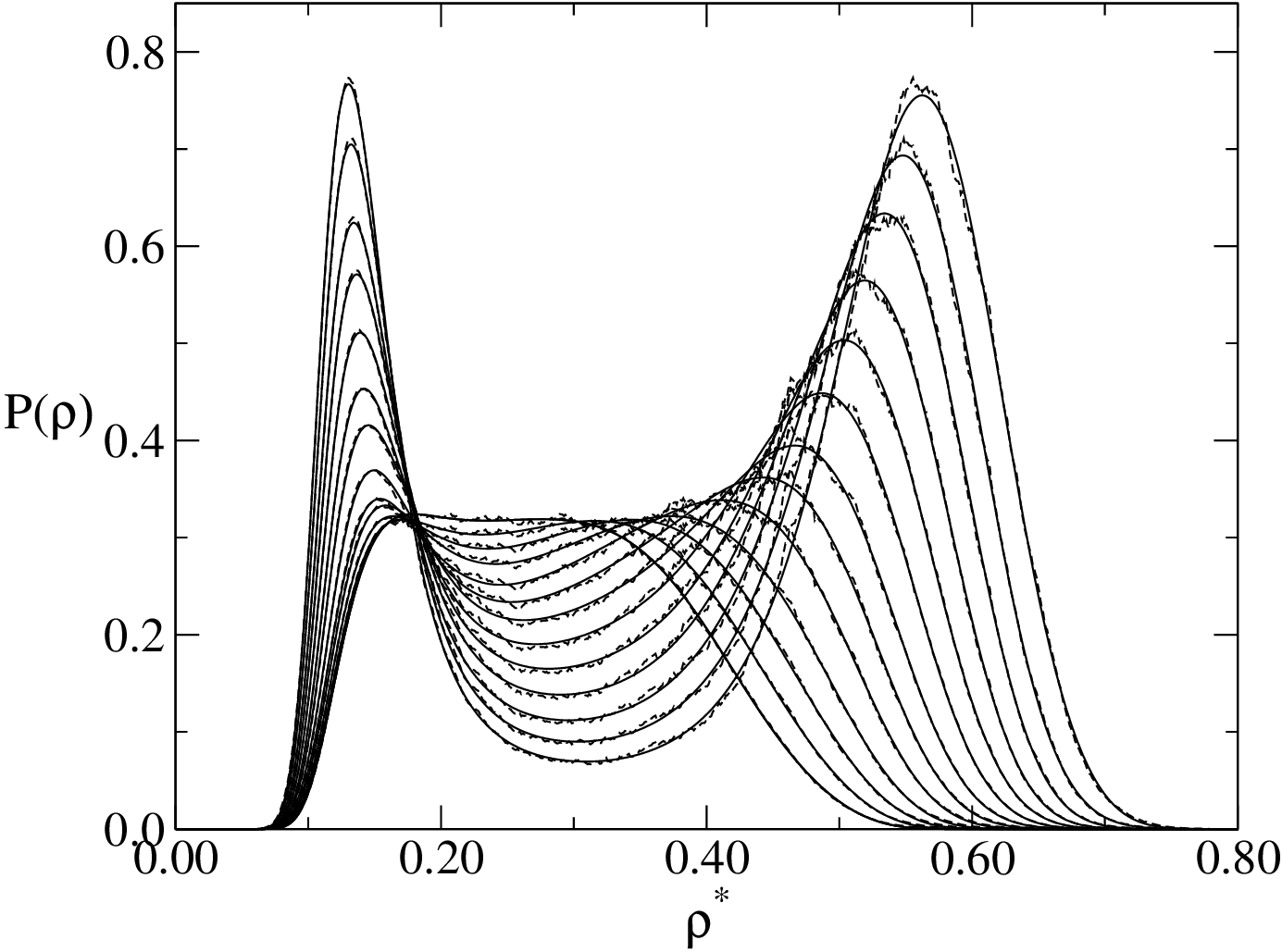}
\caption{
\label{pdro} Density histograms at coexistence for a reduced volume $V^* =2000$ and
reduced temperatures
in the range $0.0126 <T^*< T^*=0.0150$ by increments of  $\delta T^* =0.0002$. Dashed lines:
MC data after reweighting, solid lines: fits.
}
\end{figure}

During the simulation runs we recorded, at fixed $\mu$,
$\beta$, and $V$,  the joint distribution $p_{L}(\rho,u)$ of the numerical density $\rho$ and energy density
$u =U/V$ which is the  basic ingredient of our analysis of the critical properties. For each volume and temperature
typically a dozen of chemical potentials was considered in order to span a wide domain of liquid and gaseous states
near criticality. For each volume we thus considered more than a hundred of thermodynamic states. 
The acceptation rate for insertion or deletion of a pair of anion-cation was $\sim 10^{-4}$ and $\sim  0.3$ for
the displacements.
The total amount of generated configurations was about $10^{12}$
of which we kept only  one out of $1500$ to build the histograms.
Use of multi-histogram reweighting was made to infer the joint distribution $p_{L}(\rho,u)$ for a quite wide
domain in the $(\beta, \mu)$ plane, close to the critical point $(\beta_c, \mu_c)$, 
from the ones obtained for each individual state $(\beta_i, \mu_i)$ 
considered in the MC runs \cite{Ferrenberg}. 
Table~I summarizes some of the technical characteristics of our simulations. 
\section{Results}
\label{Results}

\begin{figure}[t!]
\centering
\includegraphics[angle=0,scale=0.60]{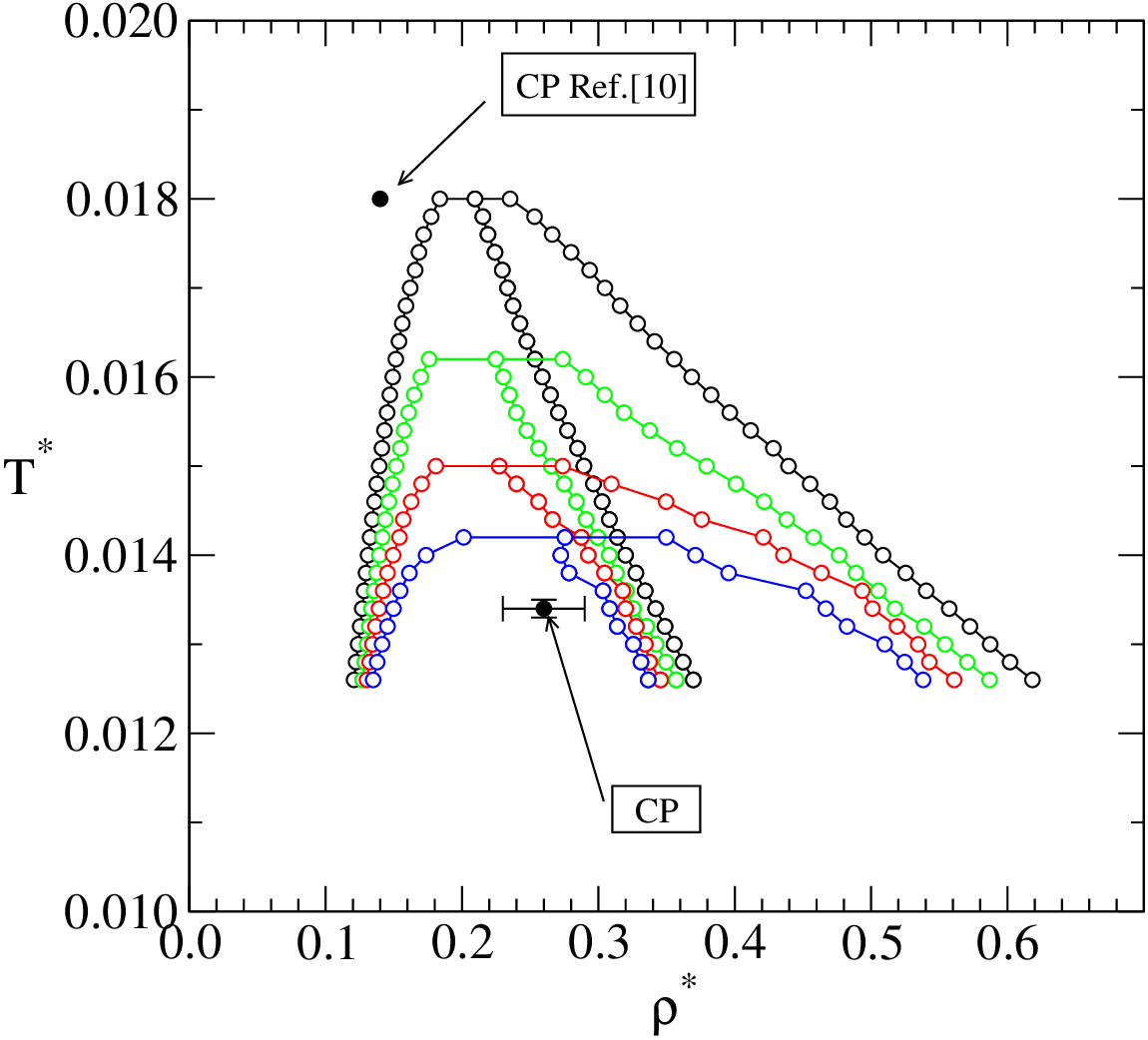}
\caption{
\label{coex} Coexistence curves $\rho_g(T^*)$ and  $\rho_l(T^*)$ for the volumes $V^*=500$ (black),  $V^*=1000$ (green),
 $V^*=2000$ (red),  and $V^*=4000$ (blue). We also display  the curves $(\rho_g(T^*) + \rho_l(T^*))/2$
which roughly satisfied the law of rectilinear diameters. Solid circles : estimates for the critical points.
}
\end{figure}

Our simulations confirm the existence of the liquid-vapor transition of the URPM
discovered in refs~\cite{Coslo-I,Coslo-II}.
For each considered volume $V$ and for each temperature smaller than some effective $T_c(L)$,  we obtained 
a bunch of bimodal density histograms $p_L(\rho)$ for chemical potentials  $\mu$  close to the chemical potential
 $\mu_{\mathrm{coex}}(T)$
at coexistence. Here $\mu_{\mathrm{coex}}(T)$ is defined as  the one that ensures that  the two peaks of the histogram
have the same height.
It then follows that  the pressures of the two coexisting phases are  equal.
Figure~\eqref{pdro} displays
some  histograms  $p_L(\rho)$ at coexistence for $T < T_c(L)$ and  the reduced volume $V^*= 2000$.

Once  $\mu_{\mathrm{coex}}(T)$ determined, $p_L(\rho)$ is conveniently  fitted by the exponential of a polynomial and
the densities $\rho_l(T)$ and $\rho_g(T)$ of the liquid and the gas are obtained  as the zeros of the derivative of this polynomial.
At some temperature $T_c(L)$ only one zero survives, giving an estimate of an apparent critical temperature $T_c(L)$.
The coexistence curves for the different volumes are given in Figure~\eqref{coex}.
Note that the law of rectilinear diameters is only but
roughly satisfied and  the critical density  $\rho_c(L) $  is obtained from an extrapolation. 

\begin{figure}[t!]
\centering
\includegraphics[angle=0,scale=0.60]{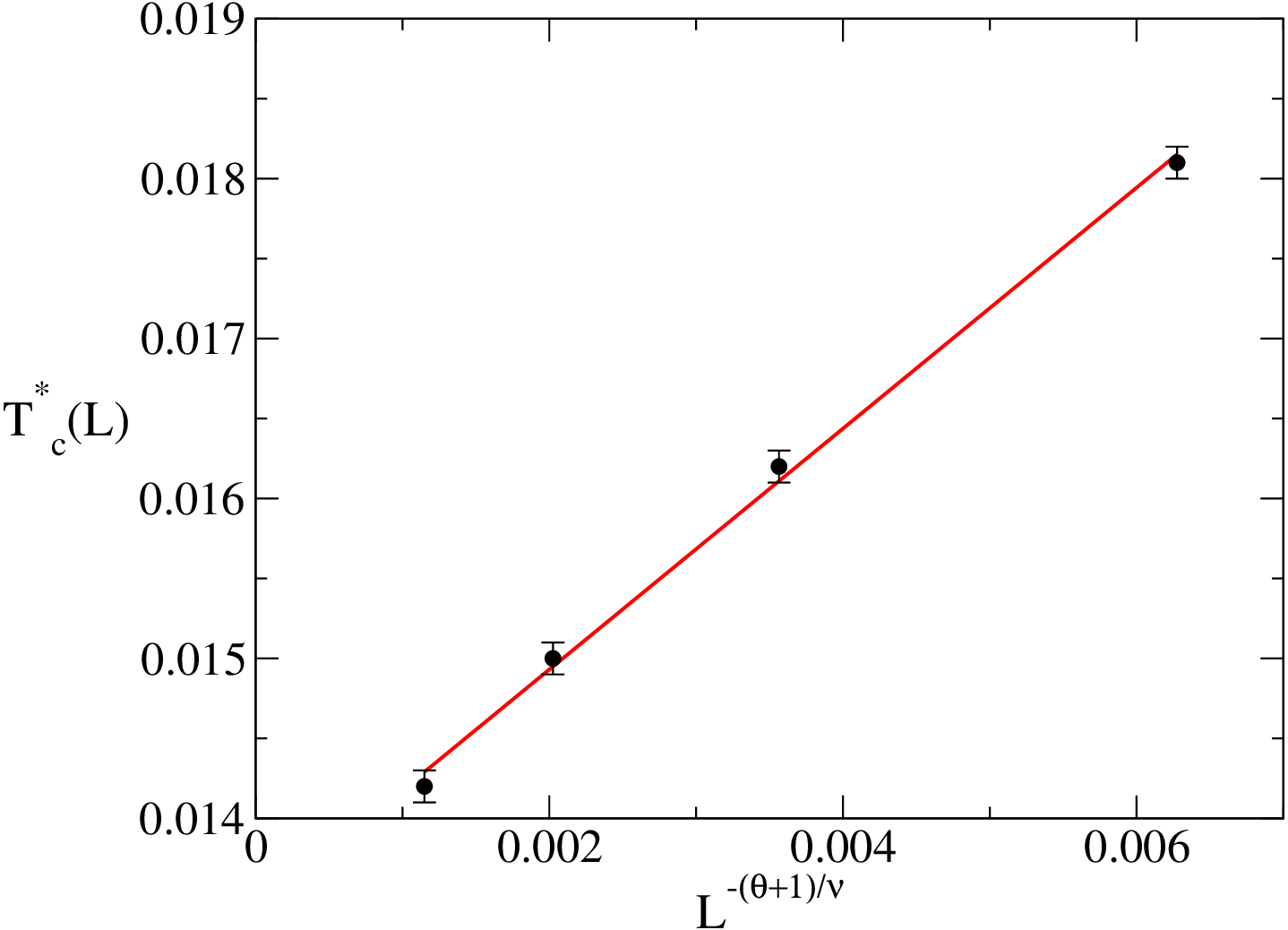}
\caption{
\label{Tc-apparent} Apparent critical temperature $T_c^*(L)$, as defined in the text, vs  $L^{-(1+\theta)/ \nu}$,
where the exponents $\theta$ and $\nu$ are those of the universality class of the 3D-Ising model .
The $T_c^*(L)$ are $0.0181$,  $0.0162$, $0.0150$, and $0.0142$ for $V^*=500$,
$V^*=1000$,  $V^*=2000$, and $V^*=4000$ respectively.
Red solid line : linear regression of the MC data \mbox{
$y=0.013424 + 0.75334*x$},  giving the infinite volume limit  $T_c^*(\infty)=0.0134(1)$.
}
\end{figure}

\begin{figure}[t!]
\centering
\includegraphics[angle=0,scale=0.60]{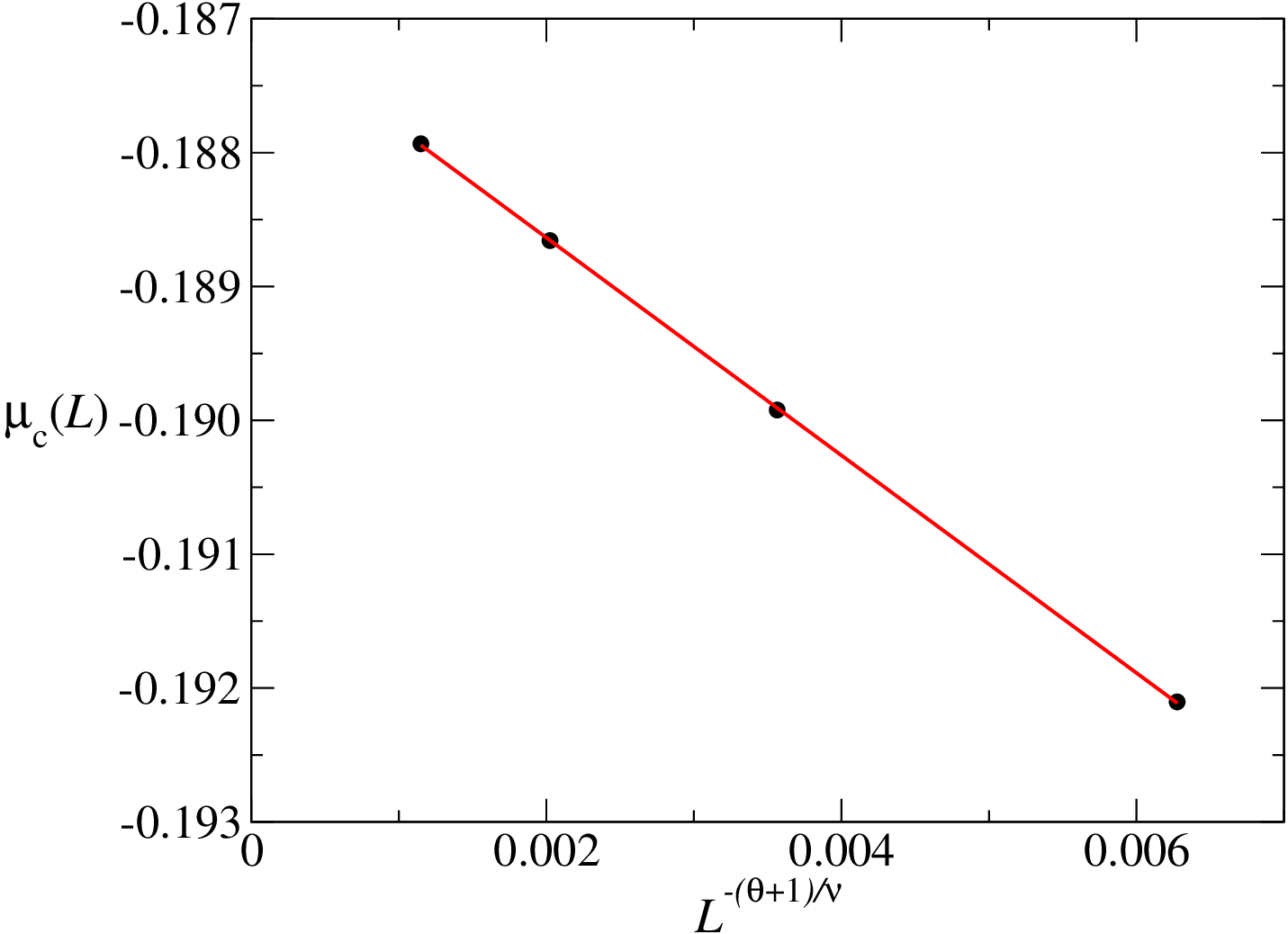}
\caption{
\label{muc-apparent} Apparent critical chemical potential $\mu_c(L)$, as defined in the text, vs  $L^{-(1+\theta)/ \nu}$,
where the exponents $\theta$ and $\nu$ are those of the universality class of the 3D-Ising model .
Red solid line : linear regression  \mbox{
$y=-0.18701 - 0.81317*x$.} The error bars on  $\mu_c(L)$ are smaller than the sizes of the symbols
}
\end{figure}

As apparent on Figure~\eqref{coex} the curves depend strongly on the system size $L$. 
Although such a huge dependency seems to imply that the considered volumes 
are not large enough to achieve the critical scaling regime, 
the $T_c(L)$  values have been used to obtain an estimate of $T_c(\infty)$
by supposing $T_c(L)$ linearly dependent on $L^{-a}$ as in the scaling limit. 
It turns out that the $a$ value is very close from the value expected for 
an Ising-like criticality, {\it i.e} $a\simeq (\theta+1)/\nu$  where $\nu$ and $\theta$
are respectively the correlation length exponent and correction to scaling
Wegner exponent~\cite{Bruce,Wilding,Wegner}. The Ising values of $\nu$ and $\theta$ are 0.63 and 0.53
and give $a \sim 2.44$.
The curve of  $T_c(L)$ versus $L^{-(\theta +1)/\nu}$ is displayed in Figure~\eqref{Tc-apparent}
and 
Table~\ref{Tab-1}.
Indeed a  typical  linear behavior is obtained leading to
an extrapolated critical temperature  $T_c^* \equiv T_c^*(\infty)=0.0134 \pm 0.0001$. 
Note that, by contrast, making the hypothesis
of a mean field criticality  $T_c(L)$ would rather scale as $L^{-3}$ (no hyperscaling!) \cite{Luiten} which, however, 
is not the   behavior that we observed.
This estimate differs from that given in Ref.~\cite{Coslo-I}, \textit{i.e.} $T_c^* = 0.018$ 
which was obtained for smaller system sizes (\textit{cf.} above)
and without finite scaling size analysis.

It turns out that the chemical potential also scales with system size as
 $\mu_c(L) \sim L^{-(\theta +1)/\nu}$ yielding
  $\mu_c =-0.18701 \pm 0.00001$, with an excellent precision,   
see Figure~\eqref{muc-apparent} and 
Table~\ref{Tab-1}.
The critical 
density in the TL  is obtained with  much less precision and one finds  $\rho_c^* =0.26 \pm 0.03$.

In addition to this heuristic and aproximate determination of the critical point
in the thermodynamic limit, we have attempted quantitative FSS analyses of 
our URPM simulation data by using several theoretical frameworks proposed to determine
the critical properties. Theses analyses confirm that these data are not in the 
scaling regime. 

\begin{itemize}
 \item{(i)} Field mixing.

Following the seminal work of Bruce and Wilding~\cite{Bruce,Wilding} many simulation results of off-lattice
critical fluids have been analyzed along the lines of the revised scaling theory of Rehr and Mermin~\cite{Rehr}.
In this (approximate) analysis one establishes  a mapping between the fluid and the 3D-Ising model which restores
the $Z2$ symmetry. The two relevant scaling operators $\mathcal{M}$ (magnetization)
and $\mathcal{E}$ (magnetic energy) of the associated Ising model are supposed to be linear
combinations of the fluid variables $\rho$ and $u$ near criticality. For a given volume $V$ and 
at a given apparent critical temperature the histogram $p(\mathcal{M})$ of the order parameter should
collapse on a universal function, typical of the 3D Ising universality class   $p^*(\mathcal{M})$
known from lattice spin simulations~\cite{Blote}.
In ref.~\cite{Caillol-I} the MC data on the RPM have been successfully analyzed within this framework.
Here, for the URPM, and quite unexpectedly, it was impossible to find reasonable  field mixing parameters 
in order obtain a collapse of  $p(\mathcal{M})$ onto the universal $p^*(\mathcal{M})$. Clearly, even
 the largest considered volume $V^*=4000$ seems far from the scaling regime.

 \item{(ii)} The locus $\chi_{N^3}=0$.

This scheme was proposed in Ref.~\cite{Orkoulas}: one studies several well chosen  density and/or energy fluctuations
\mbox{$\chi_{N^kU^m}= <(N-<N>)^k (U-<U>)^m>$} along the locus $\chi_{N^3}=0$ of the phase diagram (to which
the apparent critical point at volume $V$ necessarily belongs). This is an alternative way used to restore 
the $Z2$ symmetry in the model.
Some fluctuations exhibit extrema along this locus,
the position and the height of which should scale smoothly with system size. Our data reveal a non monotonous
behavior with $L$ which suggests that the scaling regime has not yet been reached.

 \item{(iii)} The Q-locus.

We follow here a suggestion by Kim and Fisher~\cite{Kim} very similar to the previous item. Binder's cumulant~\cite{Binder}
$Q_L=\chi_{N^2}^2/\chi_{N^4}$ is computed along the Q-locus, \textit{i.e.}~the locus of the maxima of $Q_L$ at a given $T$
in the $T, \mu$ plane. Scaling laws reminiscent to that of the Ising model then should apply to $Q_L$ along the Q-locus.
This analysis turned out not to be possible for the URPM mainly because the intersections of the 
various curves $Q_L(T)$ do not follow a monotonous behavior.
\end{itemize}

These failures strongly suggest   that our simulations are far from the scaling regime and that the usual
FSS analysis is not relevant.
It should be mentioned that the range of the apparent critical  temperatures spreads from 
$T_c^*(L) = 0.0181 \sim 1.35 \times T_c^*({\infty})$ for $V=500^*$
to $T_c^*(L) =0.0142    \sim 1.059 \times T_c^*({\infty}) $ for $V=4000^*$,  with an
infinite volume extrapolation of $T_c^*({\infty}) = 0.0134$.
These ranges of variations are huge when compared
to those considered in the simulation  of  the RPM in Ref.~\cite{Caillol-I} 
where  all the apparent critical temperatures $T_c(L)$
do not differ from   $T_c(\infty)$  by more than $0.35 \%$. Our simulations of the
URPM involved too small samples and the scaling limit was probably not yet obtained. 


\section{Conclusion}
\label{Conclusion}
We have performed MC simulations of the ultrasoft restricted primitive model of polyelectrolytes in the
grand canonical ensemble. Our simulations confirm the existence
of a liquid-vapor transition and allow us to give the following estimations for its critical point
\begin{itemize}
\item $T_c^*=0.0134(1)$ 
\item $\mu_c =-0.18701(1)$
\item $\rho^*_c=0.26(3)$
\end{itemize}
These estimates were obtained by a partial FSS analysis assuming a 3D-Ising like criticality.
A full FSS analysis of the MC data was however not possible since  the scaling limit
was not reached, even in the case of  the largest volume $V^*=4000$ considered in the study.
A complete FSS anaysis would thus require to consider much larger volumes than $V^*=4000$.
The high value of the critical density, \textit{i.e.}~$\rho^*_c=0.26$ would imply the simulation 
of liquid phases involving several thousand of polyions which seems unrealistic.

As mentioned in the introduction further studies
of Coulomb criticality are wanted in next future. We suggest to  consider the model with hard cores (\textit{i.e.}~ the RPM)
rather than the URPM to endeavor such studies.
The URPM remains however appealing to test analytical theories.

\end{document}